\documentclass{article} 
\usepackage{amsmath}
\usepackage{cases}
\usepackage{multicol} 
\usepackage{authblk}
\usepackage[a4paper,left=25mm,right=25mm,top=25mm,bottom=25mm]{geometry}

\title{The Controller Generated from Noise Can Be Lyapunov Stable: A Controller Stabilized Method}  
\author{MA Le, CUI Kaige, YAN Yiming,WU Xiaoyue, GAO Nan}   
\affil{Robotics Lab,Northeast Electric Power University,China}
\date{}

\begin{document}
\maketitle

\begin{abstract}
Aiming at the difficulty of stability analysis in practical application of existing control methods, a controller strategy based on lyapunov stability theory is proposed to realize stable control for any control method. In order to illustrate the effectiveness of this control scheme, this paper lists two different lyapunov functions to test the control effect of different system models in tracking the target function. The simulation results show that the proposed method has excellent control effect on different types of systems and is advanced.
\end{abstract}

\section{Introduction}
\par The stability of system is critical issue of controller analysis. Stability means that the motion of the system removed the disturbances can revert to the original state of motion[1], and System stable analysis is a key to ensuring the stable operation of the system. Many advanced control methods have been presented to achieve better control of the industrial process, such as proportional integral derivative (PID)[2-3], model predictive control(MPC)[4-5] and active disturbance rejection control(ADRC)[6-7]. However, all of these control methods are difficult to implement a simple proof of stability. Therefore, in order to address these issue, controller stabilization methods have to be designed.

\par Lyapunov stability theory is a very popular tool employed in stability analysis of modern control theory. The advantage of this analysis method is that it avoided the solution of complex differential equations by setting a positive definite Lyapunov function and judging the stability of the system by its differential and negative definiteness[8]. There are many control methods designed by using Lyapunov stability theory as constraints, such as adaptive control and back-stepping control. The design process for those control methods are what we define "positive" process.

\par In this paper, we present a controller $u$ using the Lyapunov stability theory and setting variable $u_{b}$ to simulated the noise signal of the controller in actual operation, which provides an effective solution for simplify of the proof of stability for us by combined the controller and control method. we can achieve "reverse" stabilization for these control method which is difficult to prove stability by using this method. Based on simulation results, the performance of the controller which this paper design can achieve the desired control effect.

\section{Lyapunov Stabilizing}
\par In this section, we use the fundamental results of Lyapunov stability theory to stabilize a control law $u_{b}$ by limiting it with dynamic upper and lower bounds. Consider the following second-order system:

\begin{equation}\label{1}
  \begin{cases}
      & \dot{x}_{1}=x_{2} \\
      & \dot{x}_{2}=f(x_{1},x_{2})+bu_b \\
      & y=x_{1}
   \end{cases}
\end{equation}

Define a Lyapunov function $V_{1}$:

\begin{equation}\label{2}
   V_{1}=\frac{1}{2}e^{2}
\end{equation}
where $e=y_{r}-y$, $y_{r}$denotes desired output, and  $y$ denotes actual output.

\par Differentiating $V_{1}$ with respect to time yields:

\begin{equation}\label{3}
   \begin{split}
      \dot{V}_{1}&=e \left ( \dot{y}_{r}-x_{2} \right ) \\
             &=e \left ( \dot{y}_{r} - \int_{0}^{t}{f(x_{1},x_{2})+bu_b \text{d} \tau}\right )\\
             &=e \left ( \dot{y}_{r} - \int_{0}^{t}{f(x_{1},x_{2})\text{d}\tau}
                                     - b\int_{0}^{{t}'}{u_b\text{d}\tau}
                                     - b\int_{{t}'}^{t}{u_b\text{d}\tau} \right )\\
             &=e \left ( \underset{\textit{M}}{\underbrace{ \dot{y}_{r} - \int_{0}^{t}{f(x_{1},x_{2})\text{d}\tau}
                                     - b\int_{0}^{{t}'}{u_b\text{d}\tau} } }
                                     - (t-{t}')bu_b(\varepsilon)  \right ),\varepsilon \in[{t}',t]
   \end{split}
\end{equation}
Assume that $u_b(\varepsilon)$ is constant on $\in[{t}',t]$. This equation can also be written as
\begin{equation}\label{4}
   \begin{split}
      \dot{V}_{1}&=\gamma e \left (\frac{M}{\gamma} - u_b(t) \right)
   \end{split}
\end{equation}
where $ \gamma = b(t-{t}') $.

Obviously, only when $e$ is contrary sign to $ \frac{M}{\gamma} - u_b(t) $, $\dot{V}_{1} \leq 0$ be guaranteed.

\begin{equation}\label{5}
   u=\left\{\begin{matrix}
   u_{b}>\frac{M}{\gamma} , e>0 \\
   u_{b}<\frac{M}{\gamma} , e<0
\end{matrix}\right.
\end{equation}

Hence for some control law $u_{b}$, according to Eq.(5) we get a Lyapunov stable controller $u$ and it can be expressed as

\begin{equation}\label{6}
   u=\left\{\begin{matrix}
     max \left( \frac{M}{\gamma},u_{b}  \right) , e>0 \\
     min \left( \frac{M}{\gamma},u_{b}  \right), e<0
\end{matrix}\right.
\end{equation}

Notice in this section there is an assumption that $u_{b}$ is constant on . And $\dot{e}$ is not involved, therefore let us consider the Lyapunov function $V_{2}$:

\begin{equation}\label{7}
   V_{2}=\frac{1}{2}s^{2}
\end{equation}

where $s=ke+\dot{e}$.

Differentiating $V_{2}$ with respect to time yields:

\begin{equation}\label{8}
   \begin{split}
      \dot{V}_{2} &= s\dot{s} \\
                  &= s \left( k\dot{e}+\ddot{e} \right ) \\
                  &= s \left(  \underset{\textit{N}}{ \underbrace { k\dot{e} + \ddot{y}_{r} -f(x_{1},x_{2})} } -bu_{b}  \right )
   \end{split}
\end{equation}

This equation can also be written as
\begin{equation}\label{9}
   \dot{V}_{2} = bs \left( \frac{N}{b} - u_{b} \right)
\end{equation}

Similarly, only when $s$ is contrary sign to $ \frac{N}{b} - u_{b} $ will $\dot{V}_{2}\leq 0$ be ensure. Notice, compared with Eq. (4), there is no constraint on $u$. It can be expressed as

\begin{equation}\label{10}
   u=\left\{\begin{matrix}
   u_{b}>\frac{N}{b} , s>0 \\
   u_{b}<\frac{N}{b} , s<0
\end{matrix}\right.
\end{equation}

Consequently, according to Eq. (10) for any control law  , we get controller   which is Lyapunov stable, and it can be expressed as

\begin{equation}\label{11}
   u=\left\{\begin{matrix}
     max \left( \frac{N}{b},u_{b}  \right), s>0 \\
     min \left( \frac{N}{b},u_{b}  \right), s<0
\end{matrix}\right.
\end{equation}

In conclusion, we obtain a Lyapunov stabilize method that are able to stabilize our control law $u_{b}$ and get a lyapunov stable controller $u$ even $u_{b}$ is a noise signal.

\section{Numberical Simulation Analysis}

This section shows the results of the carried out simulations in order to confirm the effectiveness of the suggested methods. Also, the controller is an extreme case.  Python program has been used in order to simulate the suggested methods.
%

\subsection{Test 1}
\par  Consider the following system:
\begin{equation}\label{12}
  \begin{cases}
      & \dot{x}_{1}=x_{2} \\
      & \dot{x}_{2}=-a_{1}x_{1}-a_{1}x_{1}+bu \\
      & y=x_{1}
   \end{cases}
\end{equation}

where $a_{1}=3$, $a_{2}=2$ ,$b=1$, sampling period $\triangle t=0.01s $. Let the initial states are$ [x_{1},x_{2}]=[0,0] $, and the simulation time is 60s. The controller $u_{b}$ is generated from a serial of stochastic noise signal, of which range is$[-500,500]$. The desired position signal is $ y_{r}(t)=sin(t)$.

Simulation results of the test 1 are shown in Fig.1. From Fig.1. Due to The controller $u_{b}$ is varied in the range from -500 to 500, the controller $u$ is varied in the range from -1000 to 1000. But tracking errors are varied in the range from -0.2 to 0.2. It is seen that the proposed method has a good tracking result.


\newpage
\subsection{Test 2}
The same parameters and system model as in test 1 are used with test 2. Simulation results of the test 2 are shown in Fig.2. From Fig.2. Although the desired position signal is $y_{r}(t)=1$, the system is still stable. The validity of Lyapunov function $V_{1}$ is further verified.


\subsection{Test 3}
The same parameters and system model as in test 1 and the Lyapunov function $V_{2}$ mentioned in the previous chapter are used with test 3. Simulation results of the test 3 are shown in Fig.3. It can be seen that system output is buffeting. The reason is that $u_{b}$ is generated from a serial of stochastic noise signal, which range is[-500,500]. Compared with test1, since e is constrained in Lyapunov function $V_{2}$, the range of system input u is increased to [-5000,5000]. It is seen that the proposed method has a good tracking result, even in the presence of noise.


\subsection{Test 4}
The same parameters and system model as in test 1 are used with test 4. Simulation results of the test 4 are shown in Fig.4. From Fig.4. Although the desired position signal is $y_{r}(t)=1$, The system is still stable. The validity of Lyapunov function $V_{2}$ is further verified.

In conclusion, the simulation results of test 1-4 verify the effectiveness of the proposed two methods in linear systems. The controller $u_{b}$ is generated from a serial of stochastic noise signal, It is proved that the method in this paper has some generalization for controller selection.

\section{Conclusion}
Through the above theoretical analysis and simulation, the following conclusions are drawn: 1) The simulation results show that the system has a good tracking effect. 2) The designed controller can ensure that the system outputs can accurately track the desired signals even if The control is a positive constant limited to the range. The above conclusion can be applied to the embedding of a forward controller to prove its stability.

\end{document}